\documentclass[a4paper,11pt]{article} 

\usepackage{jheppub1} 
 
\usepackage[latin3]{inputenc}
\usepackage[makeroom]{cancel}
\usepackage{graphicx}
\usepackage{amsmath}
\usepackage{amsfonts}
\usepackage{amssymb}
\usepackage{color}
\usepackage[export]{adjustbox}
\usepackage{graphicx}
\usepackage{dcolumn}
\usepackage{bm}
\usepackage{simplewick}
\usepackage{array}
\usepackage{appendix}
\usepackage{relsize}
\usepackage{subcaption}

\newcommand{\be}{\begin{equation}}
	\newcommand{\ee}{\end{equation}}
\newcommand{\beq}{\begin{equation}}
	\newcommand{\eeq}{\end{equation}}
\newcommand{\bea}{\begin{eqnarray}}
	\newcommand{\eea}{\end{eqnarray}}

\title{\boldmath Unified first law for traversable wormholes in non-minimal coupling of curvature and matter}

\author{Mudassar Rehman and}
\author{Khalid Saifullah}

\affiliation {Department of Mathematics, Quaid-i-Azam University, Islamabad, Pakistan}

\emailAdd{mrehman@math.qau.edu.pk}
\emailAdd{ksaifullah@fas.harvard.edu}

\abstract{In this paper thermodynamics of static Morris-Thorne wormholes has been discussed in the context of $f(R)$ gravity. The generalized surface gravity, unified first law of thermodynamics and wormhole dynamics have been studied at trapping horizons. We have investigated thermodynamics in non-minimal coupling of curvature and matter which produces very complex equations. Our results generalize the results that have already been derived in Einstein's gravity in the absence of curvature-matter coupling.
\vspace{85 mm}
}

\notoc

\begin{document}
\maketitle


\section{Introduction}

\label{sec:intro}

The general theory of relativity, established in 1915 by Einstein, has proven to be a successful theory both on theoretical as well as observational fronts \cite{RFN1, RFN2}. Recently this theory has successfully been confirmed by the detection of gravitational waves \cite{RFN3, RFN4}. The gravitational field equations in this theory, by adopting the gravitational Lagrangian density $\mathcal{L}_{m}=R$, are given by \cite{RFN1, RFN2, RFN5}
\begin{equation}\label{}
  R_{\mu \nu}-\frac{1}{2}Rg_{\mu \nu}=8\pi T_{\mu \nu}. 
\end{equation}
Here $R_{\mu \nu}$ is the Ricci tensor, $R$ the Ricci scalar, $T_{\mu \nu}$ the stress-energy tensor and $g_{\mu \nu}$ the metric tensor. These equations describe the gravitational phenomena of normal matter very well but the theory cannot satisfactorily explain some phenomena on large scale like the accelerated expansion of the universe, dark matter, quantum gravity and cosmic inflation. As observational data has revealed that our universe is presently undergoing accelerated expansion \cite{RFN6}, the late-time cosmic acceleration produces imbalance in gravitational field equations. This accelerated expansion is one of the major problems that Einstein's general relativity could not solve. This accelerated expansion can be explained to some extent by adopting the Lagrangian, $\mathcal{L}_{m}=R-2 \Lambda$, where $\Lambda$ is the cosmological constant. Thus Einstein's field equations become \cite{RFN1, RFN5}
\begin{equation}\label{}
   R_{\mu \nu}-\frac{1}{2}Rg_{\mu \nu}+\Lambda g_{\mu \nu}=8\pi T_{\mu \nu}.
\end{equation}
These latter equations imply that our universe expands acceleratedly. But there is still some doubt (see Refs. \cite{RFN7, RFN8, RFN9} for more details)
in the explanation of this phenomenon and the above mentioned problems as well. In order to resolve these issues several possibilities have been proposed in the literature, that range from different models of dark energy to theories of modified gravity. Einstein's field equations were first obtained by Hilbert using an action principle in which gravitational Lagrangian density was a linear function of the scalar invariant $R$. However, there is no evidence that gravitational Lagrangian density must be only a linear function of $R$. Thus a modification of Einstein-Hilbert action was proposed to explain this accelerated expansion and other problems that remained unexplained. In this modified gravitational Lagrangian density a function $f(R)$ was introduced \cite{RFN10} and later it was further investigated and developed \cite{RFN11, RFN12, RFN13}. To obtain the modified field equations in the context of $f(R)$ gravity, usually the metric approach is used in which the action is varied with respect to the metric $g_{\mu \nu}$. But there are other approaches also which are used in literature such as Palatini formalism \cite{RFN14, RFN15}, where both connections and the metric are considered as separate variables, and the metric-affine formalism \cite{RFN15}, in which matter part of the action is varied with respect to the connection. Thus, modified gravitational field equations were obtained that explained the late-time accelerated expansion of the universe and  other problems mentioned above, in the context of $f(R)$ gravity. There have been attempts to explain the galactic dynamics of massive test particles without involving dark matter in the context of $f(R)$ gravity models as well \cite{RFN16}. 

Wormholes connect two distant regions of the same universe or different universes through the throat which must be open all the time for a wormhole to be traversable. Thus wormholes are objects like a tunnel. The name wormhole was firstly suggested by Misner and Wheeler \cite{RFN17}, although it was not a new idea. In early 20th century, many authors including Flamm \cite{RFN18}, Weyl \cite{RFN19}, Einstein and Rosen \cite{RFN20} discussed these objects.  However, Morris and Thorne constructed a spherically symmetric and static wormhole in 1988 which was also traversable as it did not contain the event horizon \cite{RFN21}. After that many attempts were made to generalize the spacetime by introducing time dependent factors in the metric. Wormholes can also play the role of time machines if one of its mouth is moved relative to the other \cite{RFN22}. Wormhole spacetime structure is supported by exotic matter which violates the null energy condition (NEC) and weak energy condition (WEC), according to Einstein's field equations. A recent astronomical data \cite{RFN23} shows that a major part of our universe consists of fluid that violates NEC \cite{RFN24}. Sushkov \cite{RFN25} and Lobo \cite{RFN26} have shown independently that phantom energy is a possible candidate for exotic matter which has the property to violate NEC and it is the energy that supports the traversable wormhole spacetime. Exotic matter is also responsible for cosmic inflation at early times and accelerated expansion at late times. Einstein's general relativity could not explain this acceleration but $f(R)$ gravity does this without the presence of exotic matter. After the discovery of thermodynamics of black holes \cite{RFN27, RFN28, RFN29}, scientists believed that there could be a connection between Einstein's equations and thermodynamics. Jacobson derived Einstein's equations and the first law of thermodynamics from the proportionality condition of entropy and area of black hole horizon \cite{RFN30}. Now, as the exotic matter and ordinary matter are time-reversed versions of each other, one may also think that wormholes and black holes are also the time-reversed versions of each other if thermodynamical behaviour of both is same. These investigations have improved the physical status of wormholes \cite{RFN31, RFN32}. 

Here in this paper we will discuss thermodynamics of spherically symmetric traversable wormholes in the context of $f(R)$ gravity using Kodama vector at the trapping horizon \cite{RFN33, RFN34, RFN35, RFN36}. Trapping horizons are the surfaces that are foliated by surfaces which are marginal. In this formalism local quantities are brought into use to study the thermodynamical behaviour of spacetime instead of global quantities as global properties cannot be located by the observers. Further, here Kodama vector is used instead of Killing vector and trapping horizon is used in place of Killing horizon. The unified first law has been derived using the field equations and generalized surface gravity has been worked out. Both the results correspond to the formulations that have been constructed in the case of black holes \cite{RFN37}.

This paper has been organized as follows. In Section 2 gravitational field equations have been introduced.  Section 3 deals with the spacetime metric and Section 4 explains the formalism for finding trapping horizons. In Section 5 the generalized surface gravity has been worked out for static Morris-Thorne wormhole spacetimes. The unified first law of thermodynamics and wormhole dynamics have been derived in Sections 6 and 7, respectively, by using gravitational field equations. Finally we have concluded our work in Section 8.

\section{Gravitational field equations}

Consider the action of the non-minimal coupling between the curvature and matter in the context of $f(R)$ gravity given by \cite{RFN38}
\begin{equation}\label{b1}
  S=\int \left[ \frac{1}{2k}f_{1}(R)+\left\{1+\lambda f_{2}(R)\right\}L_{m}\right]\sqrt{-g}d^{4}x ,
\end{equation}
where $f_{i}(R)$ ($i=1,2$) are arbitrary functions which depend upon $R$, $L_{m}$ is the matter Lagrangian density and $k=8\pi$, $g$ is the determinant of the metric tensor and $\lambda$ is the coupling constant that characterizes the strength of interaction between curvature and matter. To obtain gravitational field equations we vary this action with respect to the metric $g_{\mu \nu}$ and get the following equations
\begin{align}\label{b2}
  F_{1}(R)R_{\mu \nu}-\frac{1}{2}f_{1}(R)g_{\mu \nu}&-\nabla _{\mu} \nabla _{\nu}F_{1}(R)+g_{\mu \nu}\Box
   F_{1}(R)=-2\lambda F_{2}(R)L_{m}R_{\mu \nu} \nonumber \\
   &+2\lambda \big(\nabla _{\mu} \nabla _{\nu}-g_{\mu \nu}\Box \big)L_{m}F_{2}(R)+\big[1+\lambda f_{2}(R)\big]T_{\mu \nu}^{m},
\end{align}
where we have used the notation $F_{i}(R)= \partial f_{i} / \partial R$. The matter stress-energy tensor reads 
\begin{equation}\label{b3}
  T_{\mu \nu}^{m}=-\frac{2}{\sqrt{-g}}\frac{\delta (\sqrt{-g}L_{m})}{\delta g^{\mu \nu}}.
\end{equation}
Here, for simplicity, we take perfect fluid which can be completely described by two quantities, the energy density and pressure. In components form it is given by \cite{RFN39}
\begin{equation}  \label{b4}
T^{t}_{t}=-\rho (r), \> T^{r}_{r}=p_{r}(r), \> T^{\theta}_{\theta}=T^{\phi}_{\phi}=p_{t}(r),
\end{equation}
where $\rho (r)$ is the energy density, $p_{r}(r)$ is the radial pressure and $p_{t}(r)$ is the tangential pressure. In this paper we will consider isotropic pressure for which $p_{r}(r)=p_{t}(r)=p(r)$.
 It has also been argued \cite{RFN40, RFN41, RFN42} that $L_{m}=p$ is the natural choice for perfect fluid, where $p$ is the pressure. This choice imposes vanishing of the extra force produced by the non-minimal coupling between curvature and matter \cite{RFN38}. No doubt, $L_{m}=p$ reproduces the fluid equations of state but this choice is not the only one \cite{RFN43}. There are other choices for the matter Lagrangian density as well, such as $L_{m}=- \rho$ and $L_{m}=-na$, where $\rho$ is the energy density, $n$ is the particle number density and $a$ is the physical free energy given by $a=\rho /n -Ts$, $s$ being the entropy and $T$ the temperature \cite{RFN42, RFN43, RFN44, RFN45}.
Here, in this paper, we will take $L_{m}=-\rho$ and consider the specific case $f_{i}(R)=R$ throughout this paper, thus gravitational field equations (\ref{b2}) reduce to
\begin{equation}\label{b5}
  G_{\mu \nu}=8\pi T_{\mu \nu}^{eff},
\end{equation}
where $G_{\mu \nu}=R_{\mu \nu}-\frac{1}{2}Rg_{\mu \nu}$ is the Einstein tensor and $T_{\mu \nu}^{eff}$ is called the effective stress-energy tensor 
\begin{equation}\label{b6}
  T_{\mu \nu}^{eff}=\big(1+\lambda R\big)T_{\mu \nu}^{m}+2\lambda \left[\rho R_{\mu \nu}-\big(\nabla_{\mu} \nabla_{\nu}-g_{\mu \nu}\Box \big)\rho \right].
\end{equation}

\section{Spacetime metric}

We consider a spherically symmetric, static, stable and traversable wormhole given by Morris and Thorne \cite{RFN21}. In coordinates $(t, l, \theta, \phi)$ this metric can be written as
\begin{equation}\label{f1}
  ds^{2}=-e^{2\Phi (l)}dt^{2}+dl^{2}+r^{2}(l)d\Omega^{2}.
\end{equation}
where $d\Omega^{2}=d\theta ^{2}+\sin ^{2}\theta d\phi ^{2}$ and $l$-coordinate runs from $-\infty$ to $\infty$. This wormhole solution covers two asymptotically flat regions which are joined together at $l=0$. This point $l=0$ is the location of the wormhole throat, the minimum radius of a wormhole, $r(l)=r_{0}$. Thus $-\infty <l<0$ and $0<l<\infty$ cover the two asymptotically flat regions. Also $e^{2\Phi (l)}$ must be finite everywhere and when $l\rightarrow \pm \infty$ then $r(l)/|l|\rightarrow 1$ and $e^{2\Phi (l)}\rightarrow$ constant in order to have asymptotically flat regions. In Schwarzschild coordinates metric (\ref{f1}) can be written as
\begin{equation}\label{f2}
  ds^{2}=-e^{2\Phi (r)}dt^{2}+\frac{dr^{2}}{1-\frac{b(r)}{r}}+r^{2}d\Omega^{2},
\end{equation}
where proper radial distance is transformed as
\begin{equation}\label{f3}
  l(r)=\pm \int \frac{dr^{\star}}{\sqrt{1-b(r^{\star})/r^{\star}}},
\end{equation}
and it should be finite everywhere implying $b(r)<r$. In metric (\ref{f2}), $\Phi (r)$ and $b(r)$ are called the redshift and the shape functions of a wormhole, since the first corresponds to the gravitational redshift of the universe and the latter describes the shape of a wormhole. Here $r$ is the radial coordinate. It decreases from $\infty$ to a minimum radius $r_{0}$, the throat of the wormhole where there occurs coordinate singularity $b(r_{0})=r_{0}$, then it increases from $r_{0}$ back to $\infty$. Thus both the flat regions are now represented by $r_{0}<r<\infty$. In order for a wormhole to be traversable the existence of event horizon should be prohibited which are the surfaces where $e^{2\Phi (r)}$ becomes zero. Thus $\Phi (r)$ should be finite everywhere for the prevention of the event horizon \cite{RFN21}. For a stable wormhole solution a flaring out condition $(b-b^{\prime}r)/b^{2}>0$ at or near the throat is imposed. Further, at throat $b(r_{0})=r_{0}$ and $b^{\prime}(r_{0})<1$ is also imposed to have a wormhole solution. The violation of NEC, in fact, is because of these restrictions \cite{RFN21, RFN46, RFN47}.

\section{Trapping horizon}

Here we use a formalism \cite{RFN35} that defines the properties of real black holes by employing local and physically relevant quantities. This formalism recovers the thermodynamical results of black holes when we use global considerations at event horizons in static vacuum case. Thus, this formalism generalizes the results of global considerations. In traversable wormhole, it is not possible to deduce any thermodynamical property using global considerations as there is no event horizon there. So we use local quantities to study the thermodynamical properties of wormholes using trapping horizon. These exhibit similar properties as those of a black hole. To obtain the trapping horizon from metric (\ref{f2}), we apply this formalism. Eq. (\ref{f2}) can be written as
\begin{equation}\label{e1}
  ds^{2}=2g_{+-}dx^{+}dx^{-}+r^{2}d\Omega^{2}.
\end{equation}
Here we have used the null coordinates $x^{+}=t+r_{\ast}$ and $x^{-}=t-r_{\ast}$ where $r$ and $r_{\ast}$ are related by the equation
\begin{equation}\label{e2}
  \frac{dr}{dr_{\ast}}=\sqrt{-\frac{g_{00}}{g_{rr}}}=e^{\Phi}\sqrt{1-b/r}.
\end{equation}
Here $r$ and $g_{+-}=-e^{2\Phi}/2$ are functions of the null coordinates $x^{+}$ and $x^{-}$. The null coordinates $x^{+}$ and $x^{-}$ are related to the outgoing and ingoing null normal radiations which are normal to each symmetric sphere $\partial_{\pm}=\partial /\partial x^{\pm}$. Here $r$ is the areal radius \cite{RFN35} and $d\Omega^{2}$ describes the metric on the unit 2-sphere. We define the expansions as
\begin{equation} \label{e3}
\Theta_{\pm}=\frac{2}{r}\partial_{\pm}r .
\end{equation}
A sphere is classified by the sign of $\Theta_{+}\Theta_{-}$ as follows: $\Theta_{+}\Theta_{-}>0$  means the  sphere is trapped, $\Theta_{+}\Theta_{-}<0$  means it is untrapped and $\Theta_{+}\Theta_{-}=0$ implies the sphere is marginal.

Now, a trapping horizon is defined as the surface foliated by spheres in which $\Theta_{+}\Theta_{-}=0$. In this paper, for trapping horizon we choose $\Theta_{+}|_{h}=0$ which implies $\partial_{+}r|_{h}=0$ giving  $b(r_{h})=r_{h}$. Also, on the throat, $b(r_{0})=r_{0}$, which implies that for metric (\ref{f2}) the trapping horizon and throat of the wormhole coincide, that is $r_{h}=r_{0}$.
Now, the trapping horizon is classified by the sign of $\Theta_{-}$ as follows: $\Theta_{-}<0$ means the trapping horizon is future, $\Theta_{-}=0$ means it is bifurcating and $\Theta_{-}>0$ implies the trapping horizon is past.
Furthermore, this trapping horizon is classified as follows: $\partial_{-}\Theta_{+}<0$ means the trapping horizon is outer, $\partial_{-}\Theta_{+}=0$ means it is degenerate and $\partial_{-}\Theta_{+}>0$ implies the trapping horizon is inner.

In our case $\Theta_{+}|_{h}=0$ implies $\Theta_{-}|_{h}=0$, so we have a bifurcating trapping horizon here. However this bifurcating trapping horizon may be outer, inner or degenerate depending on the sign of $\partial_{-}\Theta_{+}$.

\section{Generalized surface gravity}

In this section we will derive the expression of the generalized surface gravity in the framework of $f(R)$ gravity. The Misner-Sharp energy is the active gravitational energy in spherically symmetric spacetimes. This energy, for a perfect fluid, converts into Newtonian mass. In vacuum it yields Schwarzschild energy. It gives Arnowitt-Deser-Misner energy $E_{_{ADM}}$ and Bondi-Sachs energy $E_{BS}$, at spatial and null infinities, respectively \cite{RFN34}. The formula for Misner-Sharp energy is  \cite{RFN48}
\begin{equation}\label{g1}
E=\frac{r}{2}(1-\partial^{a}r\partial_{a}r)=\frac{r}{2}(1-2g^{+-}%
\partial_{+}r\partial_{-}r).
\end{equation}
At the trapping horizon $b(r_{0})=r_{0}$ and $E=r_{0}/2$.

Now, from the effective stress-energy tensor, on using background fluid, two invariants expressed in local coordinates can be constructed as
\begin{eqnarray} \label{20}
\omega &=&-g_{+-}T^{+-(eff)} \nonumber       \\ 
&=&\frac{\rho -p}{2}+\lambda \bigg[\big(1-\frac{b}{r}\big)\left\{(\rho +p)\Phi ^{\prime \prime}+(\rho +p)(\Phi ^{\prime})^{2}-\frac{\rho -p}{r^{2}}+\frac{2p\Phi ^{\prime}}{r}-\rho ^{\prime \prime}-\rho ^{\prime}\Phi ^{\prime}\right\}  \nonumber    \\
&+&(b-b^{\prime}r)\left\{\frac{\Phi ^{\prime}}{2r^{2}}(\rho +p)+\frac{p}{r^{3}}-\frac{\rho ^{\prime}}{2r^{2}}\right\}+\frac{\rho -p}{r^{2}}\bigg],  
\end{eqnarray}
and the vector
\begin{equation}
\psi =T^{++(eff)}\partial_{+}r\partial_{+}+T^{--(eff)}\partial_{-}r\partial_{-} .
\label{21}
\end{equation}

The gravitational field equations of interest for metric (\ref{e1}) are
\begin{align}\label{g2}
  -\partial _{\pm}\Theta _{\pm}-\frac{1}{2}\Theta _{\pm}^{2}&+\Theta _{\pm}\partial _{\pm}\log (-g_{+-})=\big(1+8\lambda e^{-2\Phi}\partial_{-} \partial_{+} \Phi +4\lambda e^{-2\Phi}\partial _{-}\Theta _{+} \nonumber    \\ 
  &+6\lambda e^{-2\Phi}\Theta _{-}\Theta _{+}+4\lambda e^{-2\Phi}\partial _{+}\Theta _{-}+\frac{2\lambda}{r^{2}}e^{-2\Phi}\big)T_{\pm \pm}^{(m)} \nonumber \\
  &+2\lambda \rho \big(-\partial _{\pm}\Theta _{\pm}+2\Theta _{\pm}\partial_{\pm} \Phi-\frac{1}{2}\Theta _{\pm}^{2}\big)-2\lambda \Delta _{\pm \pm}\rho ,
\end{align}
\begin{align}\label{g3}
  \partial _{\pm}\Theta _{\mp}+\Theta _{-}\Theta _{+}-\frac{1}{r^{2}}g_{\pm \mp}&=\big(1+8\lambda e^{-2\Phi}\partial_{-} \partial_{+} \Phi +4\lambda e^{-2\Phi}\partial _{-}\Theta _{+}+6\lambda e^{-2\Phi}\Theta _{-}\Theta _{+}  \nonumber   \\ 
  &+4\lambda e^{-2\Phi}\partial _{+}\Theta _{-}+\frac{2\lambda}{r^{2}}e^{-2\Phi}\big)T_{\pm \mp}^{(m)}+2\lambda \rho \big(-\partial _{\mp}\Theta _{\pm} \nonumber \\
  &-2\partial _{\mp}\partial _{\pm} \Phi -\frac{1}{2}\Theta _{-}\Theta _{+}\big)-2\lambda \Delta _{\pm \mp}\rho ,
\end{align}
Kodama introduced the Kodama vector \cite{RFN49}, which generalizes the Killing vector and reduces to it in vacuum stationary case. In null coordinates it can be written as
\begin{equation}  \label{g4}
K=-g^{+-}(\partial_{+}r\partial_{-}-\partial_{-}r\partial_{+}),
\end{equation}
which for metric (\ref{f2}) takes the form
\begin{equation}\label{g5}
  K_{\pm}=e^{-\Phi}\sqrt{1-\frac{b}{r}}.
\end{equation}
At the trapping horizon (throat of the wormhole), $b(r_{0})=r_{0}$, $\|K\|^{2}=0$.
This vector vanishes on the hypersurface $\partial _{+}\Theta =0$, thus giving the definition of a trapping horizon. The generalized surface gravity $\kappa$ is obtained in this formalism by replacing the role of the Killing vector with Kodama vector and Killing horizon with the trapping horizon, respectively, and satisfies the following equation \cite{RFN50} on the trapping horizon
\begin{equation}\label{g6}
K^{a}\nabla _{[b}K_{a]}=\pm \kappa K_{b},
\end{equation}
which is equivalent to
\begin{equation}\label{g7}
 \kappa =\frac{1}{2}g^{ab}\partial_{a} \partial_{b}r.
\end{equation}
On using the gravitational field equations, this gives

\begin{eqnarray} \label{8}
  \kappa &=& \frac{E}{r_{0}^{2}}-4\pi r_{0}\omega  \nonumber \\ 
   &=& 2\pi \bigg[r(\rho +p)-\lambda \{\frac{(\rho +p)\Phi ^{\prime}(b-rb^{\prime})}{r}+2\frac{(\rho +p)(b-rb^{\prime})}{r^{2}}-\frac{2(\rho +p)}{r} \nonumber  \\ 
  &+& \frac{2\rho (b-rb^{\prime})}{r^{2}}+\frac{\rho ^{\prime}(b-rb^{\prime})}{r}\}\bigg]. 
\end{eqnarray}

Note that the value of the generalized surface gravity depends on the type of trapping horizon. For the outer trapping horizon, $\partial_{-}\Theta_{+}<0$ and the generalized surface gravity is positive; for inner trapping horizon, $\partial_{-}\Theta_{+}>0$ and it is negative; and it is equal to zero for the degenerate trapping horizon, $\partial_{-}\Theta_{+}=0$.

The concept of trapping horizons has been used \cite{RFN35} to define the generalized surface gravity. Although in static Morris-Thorne spacetime, Killing vector is present, but it cannot be used to define the surface gravity as it does not vanish everywhere and thus the Killing horizon cannot be defined. However Kodama vector, which vanishes on $b(r_{0})=r_{0}$, gives the trapping horizon and thereby defines a generalized surface gravity. The Hawking temperature is defined \cite{RFN31, RFN32} as $T=-\kappa_{h}/2\pi$, which is positive, zero and negative for inner, degenerate and outer trapping horizons, respectively.

If a wormhole spacetime is characterized by the outer trapping horizon ($\partial_{-}\Theta_{+}<0$) then the temperature will be negative. Thus particles coming out of a wormhole throat are associated with negative temperature. This negative temperature is because of the phantom energy as its temperature is also negative \cite{RFN51}.

\section{Unified first law of thermodynamics}

In spherically symmetric spacetimes, one can formulate the unified first law of thermodynamics by using the gravitational field equations. According to this law the derivative of active gravitational energy, on using the gravitational equations, is divided into two terms, the work term and the energy supply term \cite{RFN35}.
In components form Eq. (\ref{21}) can be written as 
\begin{align}  \label{21b} 
\psi _{\pm}&=\pm e^{\Phi}\sqrt{1-\frac{b}{r}}\bigg[\frac{\rho +p}{4}+\lambda \big(1-\frac{b}{r}\big) \bigg\{-(\frac{\rho +p}{2})\Phi ^{\prime \prime} -(\frac{\rho +p}{2})(\Phi ^{\prime})^{2}-(\frac{\rho +p}{2r^{2}})-\frac{p\Phi ^{\prime}}{r}  \nonumber  \\ 
&-\frac{\rho ^{\prime \prime}}{2}-\frac{\rho ^{\prime}\Phi ^{\prime}}{2} \bigg\}  
+\lambda (b-b^{\prime}r)\left\{-(\rho +p)\frac{\Phi ^{\prime}}{4r^{2}}-\frac{(2\rho +p)}{2r^{3}}-\frac{\rho ^{\prime}}{4r^{2}}\right\}+\frac{\lambda}{2r^{2}}(\rho +p)\bigg]. 
\end{align} 

%

Now, taking gradient of the active gravitational energy using gravitational field equations, yields the following result 
\begin{equation}  \label{22}
\partial_{\pm}E=A\psi_{\pm}+\omega \partial_{\pm}V.
\end{equation}
This result is called the unified first law of thermodynamics \cite{RFN35}, with 

\begin{align}  \label{23}
\partial_{\pm}E &=\pm 4\pi r^{2}e^{\Phi}\sqrt{1-\frac{b}{r}}\bigg[\frac{\rho}{2}+\lambda \big(1-\frac{b}{r}\big)\left\{-\frac{\rho}{r^{2}}-\rho ^{\prime \prime}-\rho ^{\prime}\Phi ^{\prime}\right\}  \nonumber  \\ 
&+\lambda (b-b^{\prime}r)\left\{-\frac{\rho}{r^{3}}-\frac{\rho ^{\prime}}{2r^{2}}\right\}+\frac{\lambda \rho}{r^{2}}\bigg],
\end{align}
where $A=4\pi r^{2}$ and $V=4\pi r^{3}/3$, $r$ being the areal radius, are the area and areal
volume of the spheres of symmetry and the corresponding flat space,
respectively. The $\omega$ is called the energy density and $\psi$, the energy flux. 

The right hand side of Eq. (\ref{22}) consists of two terms: the first term, which is responsible for the variation of spacetime energy, is called the energy supply term, as due to the energy flux it produces the variation; and the second term, which supports the structure of spacetime, is called the work term which is carried out inside the wormhole.

\section{First law of wormhole dynamics}

The first law of wormhole dynamics is obtained by projecting Eq. (\ref{22}) along the trapping horizon. This projection yields the following equation
\begin{equation}\label{h1}
  E^{\prime}=\frac{\kappa A^{\prime}}{8\pi}+\omega V^{\prime},
\end{equation}
where, $E^{\prime}=z.\nabla E$, $A^{\prime}=z.\nabla A$ and $V^{\prime}=z.\nabla V$ with $%
z=z^{+}\partial_{+}+z^{-}\partial_{-}$ being the vector tangent to the trapping
horizon.

Eq. (\ref{h1}) includes in its expression the generalized surface gravity and the area. This expression looks the same as the first law of wormhole statics but here perturbations are replaced with the derivation along the trapping horizon. This first law of wormhole dynamics differs from the first law of wormhole statics in the aspect that here we use the definition of the generalized surface gravity defined at trapping horizon, instead of surface gravity defined at the Killing horizon used in the first law of wormhole statics.

 This expression (\ref{h1}) defines a relation between surface area and
geometric entropy as
\begin{equation}  \label{h2}
S \propto A|_{H}.
\end{equation}
 Eq. (\ref{h1}) can also be written in the form
\begin{equation}  \label{h3}
E^{\prime}=-TS^{\prime}+\omega V^{\prime},
\end{equation}
on the trapping horizon with
\begin{equation}  \label{h4}
S=\frac{A|_{H}}{4}.
\end{equation}

Eq. (\ref{h3}) contains negative sign in the first term on the right hand side. It is because the energy has been removed from the wormhole. Thus the first law of wormhole dynamics can be stated as ``the change in the active gravitational energy is equal to the energy removed from the wormhole and the work done in the wormhole''.

\section{Summary and conclusion}

In the present paper we have considered the Morris-Thorne wormhole which is a static spherically symmetric and traversable wormhole spacetime. We have applied the formalism \cite{RFN35} that uses the local quantities instead of global quantities to study the thermodynamics of spacetimes. These local quantities allow us to find the trapping horizon of a wormhole, which are the surfaces foliated by marginal spheres on which one of the null expansions becomes zero. Since Morris-Thorne wormhole is also a spherically symmetric spacetime, so we can apply these techniques to this wormhole also. We have first found the trapping horizon of the wormhole which is bifurcating trapping horizon due to staticity. In these static wormhole spacetimes Killing horizons do not exist despite of the presence of the Killing vector so the surface gravity could not be found by using the Killing vector. But still Kodama vector, the generalization of Killing vector, exists that reduces to the Killing vector if there is vacuum. This Kodama vector allows the presence of trapping horizon which can be used to derive the surface gravity. Thus the Kodama vector and trapping horizon play the role of the Killing vector and Killing horizon, respectively. Using this technique on Morris-Thorne wormhole, the expression of the generalized surface gravity has been derived. Later on this expression becomes the part of the first law of wormhole dynamics.

 Further a non-minimal coupling of curvature and matter has been considered, due to which, the gravitational field equations when written in the form of Einstein tensor replace the role of the stress-energy tensor with an effective tensor that consists of normal matter and curvature stress-energy tensors and reduces to the normal matter stress-energy tensor if the coupling constant $\lambda \rightarrow 0$. The expression of surface gravity and the first law of thermodynamics and wormhole dynamics, that we have obtained in this paper, have the same form as found in literature if the role of stress-energy tensor is replaced with the effective tensor. If $\lambda \rightarrow 0$ our results reduce to those obtained in Ref. \cite{RFN31}, which shows that our results are more general than those that exist in the literature.

The generalized surface gravity is positive or negative for outer or inner trapping horizon and zero for the degenerate trapping horizon. The unified first law of thermodynamics is obtained by taking gradient of the gravitational energy that results in two terms, on using the gravitational equations, the energy removal term, and the work term which is carried out inside the trapping horizon for the stable structure of the wormhole spacetime. The gravitational term and the work term have the same sign while the remaining third term, called the energy removal term, is negative in the case of a wormhole which means the wormhole structure is supported due to the presence of exotic matter. For black holes this term is positive, and is called the energy supply term, which implies that the black hole structure is supported by ordinary matter. This exotic matter is the ordinary matter under time inversion, thus wormhole can be thought as the time reversed version of a black hole and vice versa. But in the vacuum case that inversion of spacetimes is not possible which is the case of Schwarzschild and Einstein-Rosen bridge.

\section*{Acknowledgements}

Research grants from the Higher Education Commission of Pakistan under
its Project Nos. 20-2087 and 6151 are gratefully acknowledged.

\end{document}